\documentclass[doublecol]{epl2}
\usepackage{subfigure}
\usepackage{epsfig}
\usepackage{array}
\usepackage{amsmath}
\usepackage{amsfonts}
\usepackage{amssymb}
\usepackage{color}
\usepackage{graphicx}
\usepackage{mathrsfs}
\usepackage{multirow}
\usepackage{caption}

\title{Phase Transition of the Horava-Lifshitz AdS black holes}
\shorttitle{Phase Transition of the Horava-Lifshitz AdS black holes} 

\author{Yun-Zhi Du\inst{1,2} \and Ren Zhao\inst{2} \and Li-Chun Zhang\inst{2}}
\shortauthor{Yun-Zhi Du \etal}

\institute{
  \inst{1} Department of Physics - Datong University, Datong, Shanxi, China\\
  \inst{2} Institute of Theoretical Physics - Datong University, Datong, Shanxi, China
}
\pacs{04.70.Dy}{Quantum aspects of black holes, evaporation, thermodynamics}
\pacs{05.70.Ce}{Thermodynamic functions and equations of state}

\abstract{Some ones have showed the first-order phase transition of the Horava-Lifshitz (HL) AdS black holes has unique characters from other AdS black holes. While the coexistence zone of the first-order phase transition was not exhibited. As well known the coexistence curve of a black hole carries a lot of information about black hole, which provides a powerful diagnostic of the thermodynamic properties on black hole. We study the first-order phase transition coexistence curves of the HL AdS black holes by the Maxwell's equal-area law, and give the boundary of two-phase coexistence zone. It is very interesting that the first-order phase transition point is determined by the pressure F on the surface of the HL AdS black hole's horizon, instead of only the pressure P (or the temperature T). This unique property distinguishes the HL AdS black hole from the other AdS black hole systems. Furthermore, this black hole system have the critical curves, and on which every point stands for a critical point.}

\begin{document}
\maketitle

Recently, people pay more attention to the phase transition of Anti-de-Sitter (AdS) and de-Sitter (dS) black holes by regarding the cosmological constant of a n-dimensional AdS black hole $\Lambda=-\frac{n(n-1)}{2l^2}$ as the pressure $P=\frac{n(n-1)}{16\pi l^2}$. Especially the critical phenomena of phase transition of a AdS black hole in the P-V diagram were investigated in Refs. \cite{Kubiznak2012,Dolan2013,Gunasekaran2012,Frassino2014,
Wei2019,Sinovic2019,Altamirano2014a,Anabalon2018,Cai2015,Ge2015,Cai2016,Zhang2015,Zhao2013a,Zhao2013,Hendi2017,
Hendi2018,Hendi2017c,Hendi2017e,Hendi2017a,Hendi2016a,Hendi2016b,Hendi2016c,Hendi2015}. The research of the phase transition will not only help us to understand the nature of black holes more deeply, but also understand some phase transition behaviors in conformal field. The authors in Ref. \cite{Dehyadegari2017} found when choosing $Q^2-\Psi$ as the independent dual parameters, the charged AdS black hole has the similar phase transition to that in the van der Waals (vdW). Furthermore, people found that for the more parameters black hole, the behavior of phase transition is similar to that of vdW by adopting the different independent dual parameters \cite{Hendi2017,Zou2017a,Xu2015,Hendi2018a,Hobrecht2018,Maeda2018,Meng2018,Brihaye2018,
Dayyani2018,Ovguen2018,Bhattacharya2017a,Zhang2019,Majhi2017}. And the entanglement entropy in the AdS black holes also has the similar vdW's Phase Transition \cite{Johnson2014,Caceres2015,Nguyen2015,Zeng2016,Lan2019,
Zeng2016a,Wu2017,Liu2017}. The authors investigated the phase transition of a AdS black hole by the Maxwell's equal area law and the adoption of independent dual parameters $T-S$, and found the phase transition point of the entanglement entropy is consistent with the first-order phase transition point. Moreover there exists one kind of black holes, whose have the three-phase coexistence point with similar vdW's system \cite{Frassino2014,
Altamirano2014a,Anabalon2018,Zou2017a,Gunasekaran2012a,Fujimori2015,KordZangeneh2018,Wei2014}.

In Refs. \cite{Hennigar2017b,Xu2018,Ma2017b}, the $C_P-T$ curve near the critical point of the LoveLock AdS black holes and the Horava-Lifshitz (HL) AdS black holes is consistent with the $C_V-T$ curve of $^4He$. As what we have known, there exists the $\lambda$  phase transition as $^4He$ into the superfluid state. And the reasonable physical explanation have been given: the $\lambda$ phase transition may be a Bose condensation, and superfluid is related to the Bose body condensed at zero energy level. Therefore, the physical explanation of the similar phase transition of black holes is also an interesting issue. In order to give the corresponding physical explanation of phase transition, we should explore more thermodynamical properties of black holes deeply. The HL gravity, which is proposed by Horava, is a power-counting renormalizable gravity theory and can be regarded as an ultraviolet complete candidate for general relativity \cite{Li2014,Lu2009}. And the black hole solutions, thermodynamics and phase transitions of the HL black hole have attracted a lot of attention \cite{Cai2009,Mo2013,Cai2009a,Majhi2012,Wei2015}. People have given the condition of second-order phase transition and the critical exponents of HL AdS black holes by investigating the P-V diagram. Note that the first-order phase transition point of HL AdS black holes is a curve with a certain condition, and the $C_P-T$ curve near the second-order point is similar the $\lambda$ phase transition.

From the classification of the phase transition by Ehrenfest, we know that there are the obviously difference between the first-order phase transition and the second-order $\lambda$ one. For the liquid-gas phase transition in a vdW system, the $\lambda$ phase transition has the obvious signal: the heat capacity has a sharp increase before reaching the critical temperature, while the divergence of the first-order phase transition occurs when two phases coexist. The first-order phase transition will be transformed to the second-order one as $T\rightarrow T_c$ ($T<T_c$). For example, when the liquid-gas phase transition $T<T_c$, the entropy is discontinuity. And this discontinuity will be more and more small as $T\rightarrow T_c$, until it becomes zero for $T=T_c$ or $P=P_c$. At the same time, there is a vertical slope in the T-S diagram, that is the corresponding $\lambda$ phase transition. Therefore, there are some questions naturally: whether there is the  vdW-like first-order phase transition for the HL AdS black holes? If the answer is yes, what is the condition and the corresponding physical reason of the phase transition for the HL AdS black holes?

In this paper, the first-order phase transition of the four-dimensional HL AdS black holes thermodynamic system is explored by Maxwell's equal area law. We will exhibit the the condition of the two phases coexisting by adopting the independent dual parameters T-S and give the corresponding physical explanation. Furthermore the factors, which will affect the coexistence zone of the first-order phase transition, are also analyzed.

\section{Thermodynamic quantities of Horava-Lifshitz black holes}
\label{scheme2}
In this section, we will present the extended thermodynamics of the generalized topological HL black holes. The action of the HL gravity without the detailed-balance condition is \cite{Lu2009}:
\begin{eqnarray}
I=\int dt d^d x\left[L_0+(1-\epsilon^2)L_1+L_m\right]
\end{eqnarray}
with the Lagrangian of other matter fields $L_m$ and
\begin{eqnarray}
L_0&\!\!\!\!\!\!\!=\!\!\!\!\!\!\!&\sqrt{g}N\!\left[\!\frac{2}{k^2}(K_{ij}K^{ij}\!\!-\!\!\lambda K^2)\!\!+\!\!\frac{k^2\mu^2[(d\!\!-\!\!2)\Lambda R\!\!-\!\!d\Lambda^2]}{8(1\!\!-\!\!d\lambda)}\right],\\
L_1&\!\!\!\!\!\!=\!\!\!\!\!\!&\sqrt{g}N\!\left[\!\frac{k^2\mu^2R^2}{8(1-d\lambda)}
\left(1-\frac{d}{4}-\lambda\right)-\frac{k^2}{2\omega^4}Z_{ij}Z^{ij}\right].
\end{eqnarray}
Here $Z_{ij}=C_{ij}-\frac{\mu\omega^2}{2}R_{ij}$ with the Cotton tensor $C_{ij}$. In this theory, there are several parameters: $\epsilon,~k^2,~\lambda,~\mu,~\omega$ and $\Lambda$.

Compared with the general relativity, there are the relations for the parameters:
\begin{eqnarray}
c=\frac{k^2\mu}{4}\sqrt{\frac{\Lambda}{1-d\lambda}},~~~G=\frac{k^2c}{32\pi},~~~
\bar\Lambda=\frac{d\Lambda}{2(d-2)},
\end{eqnarray}
where $c,~G$ and $\bar\Lambda$ are Newton's constant, speed of light and consmological constant, respectively. We will fix $\lambda=1$ in the following, only for which the general relativity can be reconvered in the large distance approximation. In addition, we will only consider the general values of $\epsilon$ with the region $0\leq\epsilon^2\leq1$.

In this system there are the arbitrary dimensional topological AdS black holes with the metric \cite{Xu2018}:
\begin{eqnarray}
ds=-f(r)dt^2+f^{-1}(r)dr^2+r^2d\Omega^2_{d-1,k}
\end{eqnarray}
with
\begin{eqnarray}
f(r)&\!\!\!\!=\!\!\!\!&k+\frac{32\pi Pr^2}{(1-\epsilon^2)d(d-1)}-4r^{2-d/2}\times\nonumber\\
&&\!\!\!\!\sqrt{\frac{(d-2)MP\pi}{d(1-\epsilon^2)}+\frac{64\epsilon^2P^2\pi^2r^d}
{d^2(1-\epsilon^2)^2(d-1)^2}}.
\end{eqnarray}
Here $d\Omega^2_{d-1,k}$ denotes the line element of a (d-1)-dimensional manifold with the constant scalar curvature $(d-1)$, and $k=0,~\pm1$ indicate different topology of the spatial spaces. In AdS spacetime, the cosmological constant is introduced as the thermodynamical pressure \cite{Kubiznak2012}: $P=-\frac{\Lambda}{8\pi}$

The mass of this system reads
\begin{eqnarray}
M\!=\!\frac{64\pi Pr^d_+}{d(d\!-\!1)^2(d\!-\!2)}\!+\!\frac{1\!-\!\epsilon^2dk^2r^{d\!-\!2}_+}
{16P\pi(d\!-\!2)}\!+\!\frac{4kr^{d\!-\!2}_+}{(d\!-\!1)(d\!-\!2)},
\end{eqnarray}
where $r_+$ denotes the event horizon which the largest positive root of $f(r_+)=0$. The conjugate thermodynamic volume of pressure, the entropy and temperature are presented \cite{Hobrecht2018} in the following forms:
\begin{eqnarray}
V&\!\!\!\!=\!\!\!\!&\frac{64\pi r^d_+}{d(d-1)^2(d-2)}-\frac{(1-\epsilon^2)dk^2r^{d-4}_+}{16P^2\pi(d-2)}, \label{V}\\
S&\!\!\!\!=\!\!\!\!&\!\!\left\{\!\!\!
  \begin{array}{ll}
    4\pi r^2_+\left(1\!+\!\frac{3k(1\!-\!\epsilon^2)\ln r_+}{8\pi Pr^2_+}\right)\!+\!S_0 & d=3 \\
    \frac{16\pi r^{d\!-\!1}_+}{(d\!-\!1)^2(d\!-\!2)}\left(1\!+\!\frac{kd(d\!-\!1)^2(d\!-\!2)(1\!-\!\epsilon^2)}
    {32(d\!-\!2)(d\!-\!3)P\pi r^2_+}\!+\!S_0\right)& d\geq4
  \end{array}\right.\!\!\!\!\!\!, \label{S}\\
T&\!\!\!\!=\!\!\!\!&\frac{1}{8(d-1)\pi r_+[32\pi r^2_+P+kd(d-1)(1-\epsilon^2)]}\times\nonumber\\
&&\!\!\!\bigg\{1024P^2\pi^2r^4_++64k(d-1)(d-2)P\pi r^2_+\nonumber\\
&&+k^2d(d-1)^2(d-4)(1-\epsilon^2)\bigg\}.\label{T}
\end{eqnarray}
It is easy to check the first law of thermodynamics as
\begin{eqnarray}
dM=TdS+VdP+\Psi d\epsilon^2
\end{eqnarray}
with the potential
\begin{eqnarray}
\Psi=-\frac{dk^2r^{d-4}_+}{16P\pi(d-2)}.
\end{eqnarray}

\section{The construction of the Equal-Area Law in $T-S$ diagram}
\label{scheme3}
\begin{figure*}[htb]
\begin{center}
\includegraphics[width=9cm,height=8.5cm]{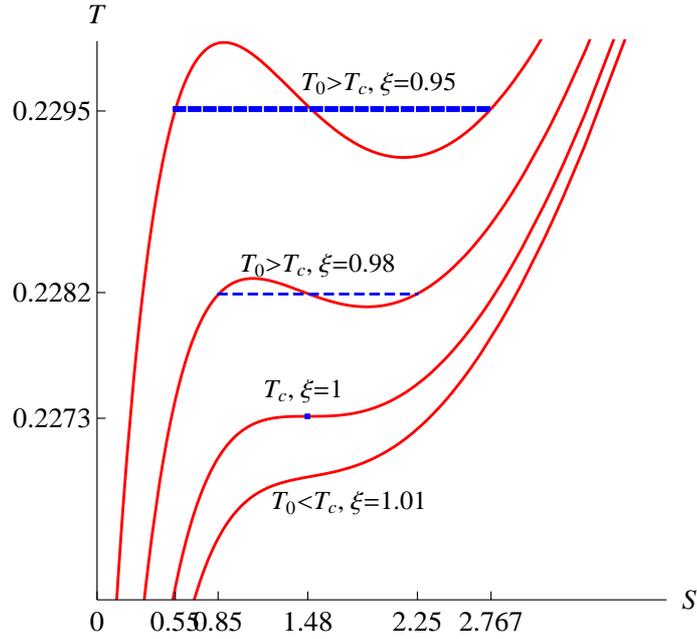}
\end{center}
\vskip -4mm \caption{The $T-S$ curves of the Horava-lifshitz AdS black hole with the different parameter values $\xi=\frac{1-\epsilon^2}{1-\epsilon_c^2}$. We set the pressure $P=0.1$ and $k=1$.}\label{k1TS}
\end{figure*}

For the HL AdS black hole thermodynamic system with the unchanged pressure in the equilibrium state, the entropies at the boundary of the two-phase coexistence area are $S_1$ and $S_2$, respectively. And the corresponding temperature is $T_0$, which is less than the critical temperature $T_c$ and is determined by the horizon radius $r_+$. Therefore, from the Maxwell's equal-area law $T_0(S_2-S_1)=\int^{S_2}_{S_1}TdS$, we have in the four-dimensional spacetime ($d=3$)
\begin{eqnarray}
0&=&T_0\left(4\pi r^2_2(1-x^2)-\frac{3k(1-\epsilon^2)\ln{x}}{2P}\right)-2kr_2(1-x)\nonumber\\
&&-
\frac{16}{3}P\pi r^3_2(1-x^3)+\frac{3k^2(1-\epsilon^2)(1-x)}{16\pi Pr_2x}\label{T0P0}
\end{eqnarray}
with $x=\frac{r_1}{r_2}$. From the equation (\ref{T}), there are the following expresses:
\begin{eqnarray}
2T_0&\!\!\!\!\!\!=\!\!\!\!\!\!&6k\epsilon^2P\!\left(\!\frac{r_1}{16\pi Pr_1^2\!+\!3k(1\!-\!\epsilon^2)}\!+\!\frac{r_2}{16\pi Pr_2^2\!+\!3k(1\!-\!\epsilon^2)}\!\right)\!\nonumber\\
&&\!\!+2(r_1+r_2)P-\frac{k}{8\pi}(1/r_1+1/r_2),\\
0&\!\!\!\!\!\!=\!\!\!\!\!\!&6k\epsilon^2P\!\left(\!\frac{r_1}{16\pi Pr_1^2\!+\!3k(1\!-\!\epsilon^2)}\!-\!\frac{r_2}{16\pi Pr_2^2\!+\!3k(1\!-\!\epsilon^2)}\!\right)\!\nonumber\\
&&\!\!+2(r_1-r_2)P-\frac{k}{8\pi}(1/r_1-1/r_2).
\end{eqnarray}
For the simplicity we define the parameter as $y\equiv\frac{16\pi r^2_+}{k}P$, so there are \begin{eqnarray}
y_1\equiv\frac{16\pi r^2_1}{k}P,~~~~~~
y_2\equiv\frac{16\pi r^2_2}{k}P\label{y22}
\end{eqnarray}
at the boundary of the two-phase coexistence area. The above equations can be rewritten as
\begin{eqnarray}
T_0&\!\!\!\!\!\!\!\!\!\!\!\!\!=\!\!\!\!\!\!\!\!&\!\!\!\!\!\frac{y_2k}{16\pi r_2}\times\nonumber\\
&&\!\!\!\!\left[\!\frac{[\!1\!+\!x]\![y_2x\!-\!\!1]}{y_2x}\!+\!\frac{\!3\epsilon^2x\!}
{\!y_2x^2\!+\!3(1\!-\!\epsilon^2)\!}\!+\!\frac{\!3\epsilon^2\!}{\!y_2\!+\!3(1\!-\!\epsilon^2)}\right],
~~~~~\label{T0}\\
0&\!\!\!\!\!\!\!\!\!\!\!\!=\!\!\!\!\!\!\!\!&\!\!\!\!y_2^3x^3+y_2^2x\left[3(1+x^2)(1-\epsilon^2)
+x(1-3\epsilon^2)\right]\nonumber\\
&&+3y_2(1-\epsilon^2)[1+3x+x^2]+9(1-\epsilon^2)^2,\label{y2}
\end{eqnarray}
and the equation (\ref{T0P0}) becomes
\begin{eqnarray}
&&\frac{4\pi r_2 T_0}{k}\left(1+x-\frac{6(1-\epsilon^2)\ln{x}}{y_2(1-x)}\right)\nonumber\\
&&=\frac{y_2(1+x+x^2)}{3}+2-\frac{3(1-\epsilon^2)}{y_2x}.\label{T00}
\end{eqnarray}
From the equations (\ref{S}) and (\ref{T}), we can obtain the critical parameters as
\begin{eqnarray}
y_c=\frac{2\sqrt{3}-1}{3},~~~~~\epsilon_c^2=\frac{4}{9}\left(1+\frac{2}{\sqrt{3}}\right).
\end{eqnarray}
For the critical point ($x=1$), substituting $y=y_c$ and $\epsilon=\epsilon_c$ into the equation (\ref{T}), the critical temperature ($T_c$), pressure ($P_c$) and horizon radius ($r_c$) satisfy the following expressions
\begin{eqnarray}
8\pi r_cT_c&=&\frac{4\sqrt{3}k}{3}, \label{Tc}\\
16\pi r_c^2P_c&=&k y_c=\frac{(2\sqrt{3}-1)k}{3}.\label{Pc}
\end{eqnarray}
For similarity, we introduce the new parameter $\xi=\frac{1-\epsilon^2}{1-\epsilon_c^2}$, the equation (\ref{y2}) can be transformed into
\begin{eqnarray}
0&\!\!\!\!\!=\!\!\!\!\!&\frac{3}{11}(15\!+\!8\sqrt{3})x^3y_2^3\!+\!y_2\xi(1\!+\!3x\!+\!x^2)\!
+\!\frac{1}{9}(15\!-\!8\sqrt{3})\xi^2\nonumber\\
&&+xy_2^2\left[(1+x+x^2)\xi-\frac{6}{11}(15+8\sqrt{3})x\right]\label{y2x1}
\end{eqnarray}
Combining the equations (\ref{T0}) and (\ref{T00}), the expression related with $y_2$ and x reads
\begin{eqnarray}
\!\!&&\!\!\!\!\!\frac{[1\!-\!x]\left(3(1+x+x^2)x y_2^2+18x y_2-(15-8\sqrt{3})\xi\right)}{x y_2[9(1-x^2)y_2-2(15-8\sqrt{3})\xi\ln{x}]}\nonumber\\
\!\!&\!\!\!\!\!=\!\!\!\!\!&\!\!\frac{\!(\!1\!+\!x\!)\!(\!x\!y_2\!-\!1\!)\!}{\!x y_2\!}\!+\!\frac{\!27x\!-\!(\!15\!-\!8\sqrt{3}\!)\!\xi x\!}{\!9x^2 y_2\!+\!(\!15\!-\!8\sqrt{3}\!)\!\xi\!}\!+\!\frac{27\!-\!(\!15\!-\!8\sqrt{3}\!)\!\xi \!}{\!9y_2\!+\!(\!15\!-\!8\sqrt{3}\!)\!\xi}.~~~~~\label{y2x2}
\end{eqnarray} It is obviously that for the given parameters $\xi$ and k, the solutions of $y_2$ and $x$ can be obtained by solving equations (\ref{y2x1}) and (\ref{y2x2}). Thus for the first-order phase transition of the HL AdS black hole with the given pressure, we can obtain the phase transition temperature $T_0$ and the horizons ($r_1$ and $r_2$) of the black hole in two different phases by substituting the values of $\xi$ and k into equations (\ref{y22}) and (\ref{T00}). The phase transition curves $T-S$ of the HL AdS black hole with different parameter values $\xi$ and the unchanged pressure $P=0.1$ are shown in Fig. \ref{k1TS}.

From Fig. \ref{k1TS} we know that the phase transition of the HL AdS black hole will appear when the redefined parameter $\xi$ is less than one (namely, $\epsilon>\epsilon_c$), and the temperature $T_0$ is bigger than the critical one $T_c$. These behaviors are fully different from the other AdS black holes \cite{Kubiznak2012,Anabalon2018,Zhang2015,Zhao2013,Hendi2017,Dehyadegari2017,
Zou2017a,Bhattacharya2017a,Majhi2017,Nguyen2015,Mo2013}. That maybe mean this system has the fully new structure from others and the physical mechanism of the phase transition is also unique.

For the first-order phase transition point of the HL AdS black hole with the given parameters ($k,~\xi$), because of $\frac{y_1}{y_2}=x^2$, there is $y=y_2$, or $y=y_1$, namely
\begin{eqnarray}
16\pi r_2^2P=ky_2=kF_2,~~\text{or}~~16\pi r_1^2P=ky_1=kF_1.
\end{eqnarray}
That indicates the first-order phase transition point is related with the horizon radius and pressure. As what we have known the pressure $F$ on the surface of the black hole's horizon reads: $F=AP=16\pi r_+^2P=kP$. Thus the defined parameter $y$ stands for the pressure $F$ on the surface of the black hole's horizon. In the other words, the first-order phase transition point is only determined by the pressure $F$ on the surface of the black hole's horizon, that is different from the charged AdS black holes \cite{Kubiznak2012,Wei2015}. When the pressure $F$ on the surface of the black hole's horizon satisfies $16\pi r_c^2P=ky_c$, the difference of the phase transition between this system and other charged AdS black holes will disappear.

From the equations (\ref{Tc}) and (\ref{Pc}), we can see that the critical temperature $T_c$ and pressure $P_c$ are not unique and are both related with the critical horizon radius. That means the parameters influencing the phase transition of the HL AdS black hole are different from other normal thermodynamic systems. With the above analyze, we find for the different values of $\xi$ in the HL AdS black hole with any given pressure, there may be a first-order phase transition, or a second-order, or nothing. Note that for the HL AdS black hole with a fixed parameter value ($\epsilon>\epsilon_c$), there is a critical curve of phase transition, not only is a critical point.

\section{Discussions and conclusions}
\label{scheme4}
In this paper we mainly study the first-order phase transition of the HL AdS black hole by the construction of the equal-area law in $T-S$ diagram. With the above analyze, the characteristics of the thermodynamic property for the HL AdS black hole are summarized as: i) It is easy to see that from the equation (\ref{V}) in the four-dimensional spacetime, the thermodynamical volume is zero when $y^2=9(1-\epsilon^2)$, while the horizon radius is not zero. That indicates there is the minimal horizon, which is related with the parameter $\epsilon$ and pressure P ($16\pi P r_{min}^2/k=3(1-\epsilon^2)^{1/2}$). ii) Since the location of horizon is independent with the temperature from  equations (\ref{y2x1}) and (\ref{y2x2}), the temperature is not the only factor to determine the phase transition, which is different from the other AdS black holes. iii) The phase transition is related with the pressure F on the surface of the black hole's horizon. In other words the phase transition of the four-dimensional HL AdS black hole with other fixed parameters ($k,~\epsilon$) is only determined by the pressure F on the surface of horizon.

\acknowledgments{
We would like to thank Prof. Zong-Hong Zhu and Meng-Sen Ma for their indispensable discussions and comments. This work was supported by the Natural Science Foundation of China (Grant No. 11475108 and Grant No. 11705106).}

\end{document}